\documentclass{natureprintstyle}
\usepackage[utf8]{inputenc}
\usepackage{units}
\usepackage{amsmath}
\usepackage{amssymb}
\usepackage{graphicx}
\usepackage{xcolor}

\makeatletter
\@ifundefined{date}{}{\date{}}
\usepackage{longtable}
\usepackage{supertabular}

\pdfoutput=1

\def\beq{\begin{equation}}
\def\eeq{\end{equation}}

\def\SI{Methods section}
\DeclareMathSizes{1}{1}{1}{1}

\makeatother

\begin{document}

\global\long\def\Mbh{M_{\bullet}}
\global\long\def\Ms{M_{\star}}
\global\long\def\Rs{R_{\star}}
\global\long\def\Ns{N_{\star}}
\global\long\def\ns{n_{\star}}
\global\long\def\Mo{M_{\odot}}
\global\long\def\Ro{R_{\odot}}
\global\long\def\Mc{M_{c}}
\global\long\def\Rc{R_{c}}
\global\long\def\Nc{N_{c}}
\global\long\def\Msig{M_{\bullet}/\sigma_{\star}}
\global\long\def\Mzo{{\mathcal{M}}_{0}}
\global\long\def\Szo{{\mathcal{S}}_{0}}
\global\long\def\Mz{{\mathcal{M}}_{z}}
\global\long\def\Gs{\Gamma_{\star}}
\global\long\def\ss{\sigma_{\star}}
\global\long\def\rh{r_\mathrm{h}}

\title{A universal minimal mass scale for present-day central black holes}

\author{Tal Alexander\thanks{Department of Particle Physics \& Astrophysics,
    Weizmann Institute of Science, Rehovot 76100, Israel.}\, and Ben
  Bar-Or\thanks{Institute for Advanced Study, Einstein Drive, Princeton, NJ
    08540, USA}}

\maketitle
\maketitle
\begin{abstract}
  Intermediate-mass black holes (IMBHs) of mass
  ${\Mbh \approx 10^{2} \textendash 10^{5}}$ solar masses, $M_{\odot}$, are the
  long-sought missing link\cite{mil+04b} between stellar black holes, born of
  supernovae\cite{abb+16a}, and massive black holes\cite{gra16}, tied to galaxy
  evolution by the empirical $\Msig$ correlation\cite{fer+00,geb+00}. We
  show that low-mass black hole seeds that accrete stars from locally dense
  environments in galaxies following a universal $\Msig$
  relation\cite{mcc+13,bos16} grow over the age of the Universe to be above
  ${\Mzo\approx3\times10^{5}\Mo}$ ($5\%$ lower limit), independent of the
  unknown seed masses and formation processes. The mass $\Mzo$ depends weakly
  on the uncertain formation redshift, and sets a universal minimal mass scale
  for present-day black holes. This can explain why no IMBHs have yet been
  found\cite{gra16}, and it implies that present-day galaxies with
  ${\ss<\Szo\approx40\,\mathrm{km\,s}^{-1}}$ lack a central black hole, or
  formed it only recently. A dearth of IMBHs at low redshifts has observable
  implications for tidal disruptions\cite{ree88} and gravitational wave
  mergers\cite{ama+13}.
\end{abstract}
The early stages of massive black hole growth are poorly
understood\cite{vol12}. High-luminosity active galactic nuclei at very high
redshift\cite{mor+11} $z$ further imply rapid growth soon after the Big
Bang. Suggested formation mechanisms typically rely on the extreme conditions
found in the early Universe (very low metallicity, very high gas or star
density). It is therefore plausible that these black hole seeds were formed in
dense environments, at least a Hubble time ago (${z\!>\!1.8}$ for a
\mbox{look-back} time of ${t_{H}\!=\!10}$ Gyr)\cite{ben+14}.

The relation ${M_{\bullet}=M_{s}{(\sigma_{\star}/\sigma_{s})}^{\beta}}$ between
black hole mass, $\Mbh$, and stellar velocity dispersion, $\ss$, that is
observed in the local Universe over more than about three decades in massive
black hole mass, correlates $\Mbh$ and $\ss$ on scales that are well outside
the massive black hole's radius of dynamical influence\cite{gra16},
${\rh\approx G\Mbh/\ss^{2}}$. Recent analyses of large heterogeneous galaxy
samples find that a universal $\Msig$ relation holds for all galaxy
types\cite{mcc+13,bos16}, although the scope of this relation and its evolution
with redshift remain contro\-versial\cite{kor+13}. Here we adopt the empirical
fit\cite{bos16}
{\small$\log_{10}(\!\Mbh/\Mo\!)\!=\!8.32\!\pm\!0.04\!\pm\!\delta_{\epsilon}\!\!+\!\!(5.35\!\pm\!0.23)\!\log_{10}(\!\ss\!/200\,\mathrm{km\,s}^{-1}\!)$},
where ${\delta_{\epsilon}=0.49\pm0.03}$ is the root mean square of the
intrinsic scatter. We assume that this universal $\Msig$ holds at all
redshifts\cite{she+15}, and that the black hole seeds grow in a locally (within
a few $\rh$) dense stellar environment. By fixing $\rh$, the $\Msig$ relation
then imposes tight connections between the black hole and the dynamical
properties of its stellar surroundings\cite{ale11}, and specifically the rate
at which it consumes stars (see \SI).

A central black hole grows by (1) the accretion of stars, compact
remnants and dark matter particles that are deflected toward it on
nearly radial orbits, and either fall whole through the event horizon
or are tidally disrupted outside it, and then accreted; (2) viscosity-driven
accretion of interstellar gas; and (3) mergers with other black holes.
Of these growth channels, only the accretion of stars must follow
from the existence of a central black hole in a stellar system. Moreover,
the tidal disruption event (TDE) rate in steady-state can be estimated
from first principles, for given boundary conditions at $\rh$~(ref.~\cite{bar+16}).

\begin{figure}[t]
  \includegraphics[width=1\columnwidth]{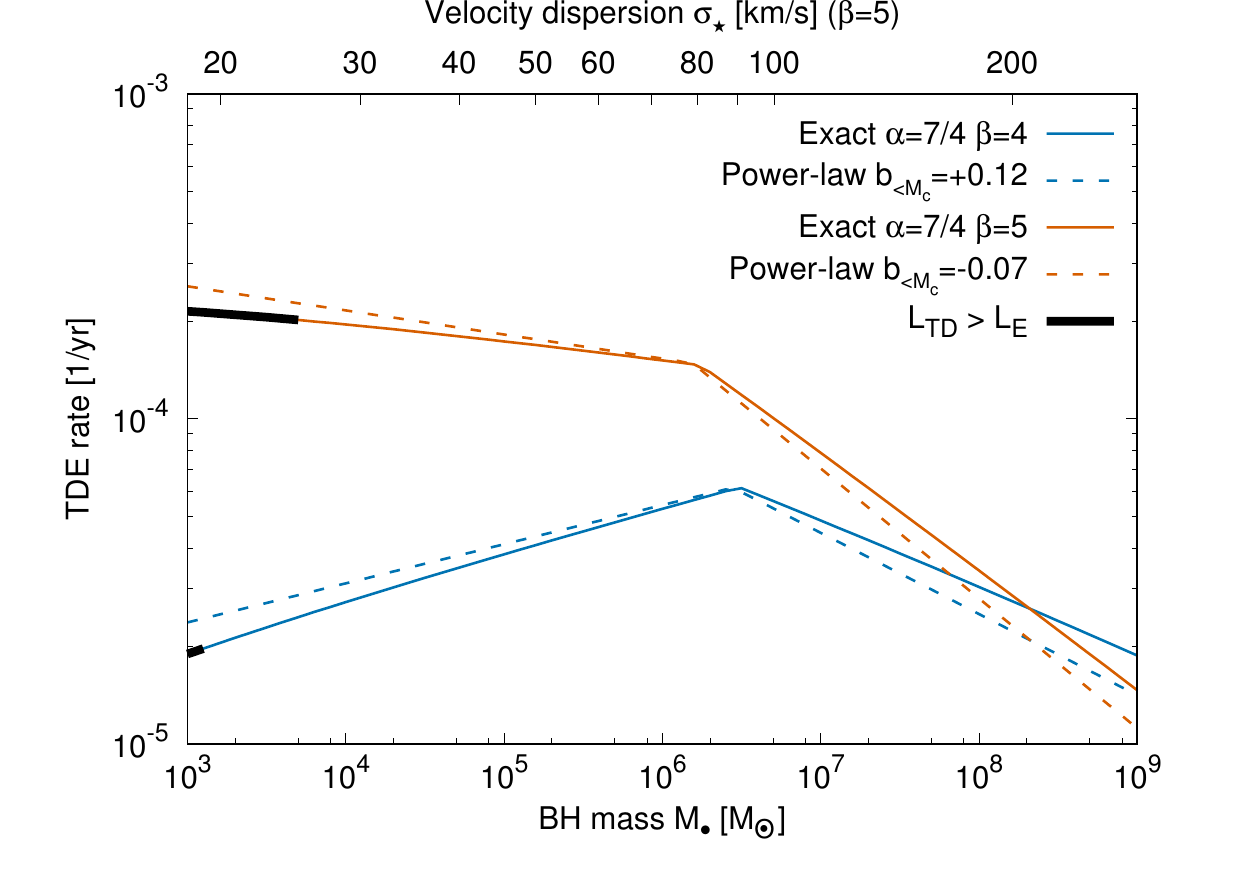}\caption{\label{f:Rprc}\textbf{Plunge
      rates as function of the black hole mass }$\protect\Mbh$\textbf{.  }Rates
    (solid lines, see \SI) are plotted for ${\Ms=1\,\Mo}$, ${\Rs=1\,\Ro}$,
    ${f_{a}=3/8}$, ${\alpha=7/4}$, and ${\beta=4,5}$, normalized to the
    empirical $\Msig$ mass parameter ${M_{s}=2.1\times10^{8}\Mo}$
    \protect\cite{bos16}. The corresponding velocity dispersion
    $\sigma_{\star}$ is displayed for a ${\beta=5}$ $\Msig$ relation. The rates
    are well-approximated by power-laws (dashed lines,
    equation~\eqref{e:TDEratePL}). Radiation back-reaction above the Eddington
    limit $L_{E}$ by the mean accretion luminosity
    ${L_{TD}=\eta f_{a} \Ms c^{2}\Gamma}$ (${\eta=0.1}$ assumed for the
    radiative efficiency) is only relevant for the lowest black hole masses,
    where it may slow down the initial black hole growth.}
\end{figure}

It has been noted that typical steady-state TDE rates, $\Gamma$,
around ${10^{-4}\,\mathrm{yr}^{-1}}$ (Fig.~\ref{f:Rprc}), imply by
simple dimensional analysis that massive black holes (MBHs) with low
mass, ${\lesssim10^{7}M_{\odot}}$, may acquire a substantial fraction
of their mass from TDEs over the Hubble time $t_{\mathrm{H}}$, or
equivalently, that linear growth by TDEs has a typical mass scale\cite{mag+99,mer13,sto+16b},
${\Mbh^{\mathrm{TDE}}\sim\Ms\Gamma t_{\mathrm{H}}\sim10^{6}\Mbh}$ (however,
the growth equation is generally nonlinear, and therefore $\Mbh^{\mathrm{TDE}}$
can significantly mis-estimate $\Mbh(t_{H})$; see \SI). Previous
studies have usually focused on the rates and prospects of TDE detection,
and not on black hole growth. Although it was recently argued that $\Mbh^{\mathrm{TDE}}$
arises as a minimal black hole mass in a specific formation scenario\cite{sto+16b},
the commonly held assumption remains that IMBHs with ${\Mbh\ll\Mbh^{\mathrm{TDE}}}$
do exist, and that this must constrain formation scenarios, or set
an upper bound on the efficiency of TDE accretion, rather than a lower
bound on IMBH masses\cite{mer13}.

Here, we argue that IMBHs are transient objects, which no longer exist in the
present-day Universe. We derive a universal lower bound on the present-day mass
scale of central black holes, $\Mzo$, that follows directly from the universal
$\Msig$ relation, and is independent of the unknown seed masses and their
formation processes. We use the $\Msig$ relation to set the boundary
conditions, and show that the nonlinear growth equation for black holes
can be bounded from below by a simple inequality that includes only growth by
TDEs. We translate the intrinsic scatter in the $\Msig$ relation to a
probability distribution for the lower bounds $\Mzo$ and $\Szo$, and show that
$\Mzo$ lies just below the lightest MBHs yet discovered\cite{gra16},
${\Mzo\lesssim\min(\Mbh^{\mathrm{obs}})\sim10^{6}\Mo}$.

Stars around a central black hole are constantly scattered in angular momentum
to nearly radial orbits below a critical (``loss-cone'') value,
${j_{lc}=\sqrt{1-e^{2}}}$ ($e$ is the orbital eccentricity), which approach the
black hole closer than the tidal disruption radius,
${r_{t}\simeq{(\Mbh/\Ms)}^{1/3}\Rs}$ ($\Ms$ and $\Rs$ are the stellar mass and
radius), where they are destroyed. Main sequence stars are disrupted outside an
IMBH's event horizon, and a fraction $f_{a}$ of about $1/4$ to $1/2$ of their
mass is ultimately accreted by the black hole\cite{aya+00}. The TDE rate
depends on the number of stars near the black hole and on the competition
between the two-body relaxation time $T_{R}$ (equation~\eqref{e:TR}) and the
orbital time in supplying and draining loss-cone orbits. The integrated
contribution in steady-state from all radii is a function of $\Mbh$ and of the
boundary conditions at $\rh$, fixed by $\ss$. The TDE rate is well-approximated
by a power-law $\Gamma\simeq\Gamma_{\star}{(\Mbh/\Ms)}^{b}$, whose index $b$ is
a function of the $\Msig$ index $\beta$, and changes across a critical mass
scale $M_{c}\sim10^{6}\Mo$ (Fig.~\ref{f:Rprc}; see \SI). 
The index ${b\ll1}$ for the empirical range ${4\lesssim\beta<6}$~(refs~\cite{gra16,kor+13}).

Let us assume that a black hole seed forms with an initial mass $M_{i}$ large enough
to dominate its radius of influence, in a central stellar system that
is massive enough to allow it to grow: that is, ${M_{\mathrm{sys}}\!\gg\! \rh^{3}\Ms\ns(\rh)\!>\!\Mbh\!\gg\!\Ms}$
at all times ($\ns$ is the stellar density). Consider first the case
where the black hole grows only by accreting stars. The black hole growth equation
is 
\begin{equation}
\dot{\Mbh}=f_{a}\Ms\Gamma_{\star}{(\Mbh/\Ms)}^{b}\equiv\dot{M}_{\bullet}^{\star}\,,\,\,\,\,\,\,\Mbh(0)=M_{i}\,.\label{e:dMbhdt}
\end{equation}

\begin{figure}
\includegraphics[width=1\columnwidth]{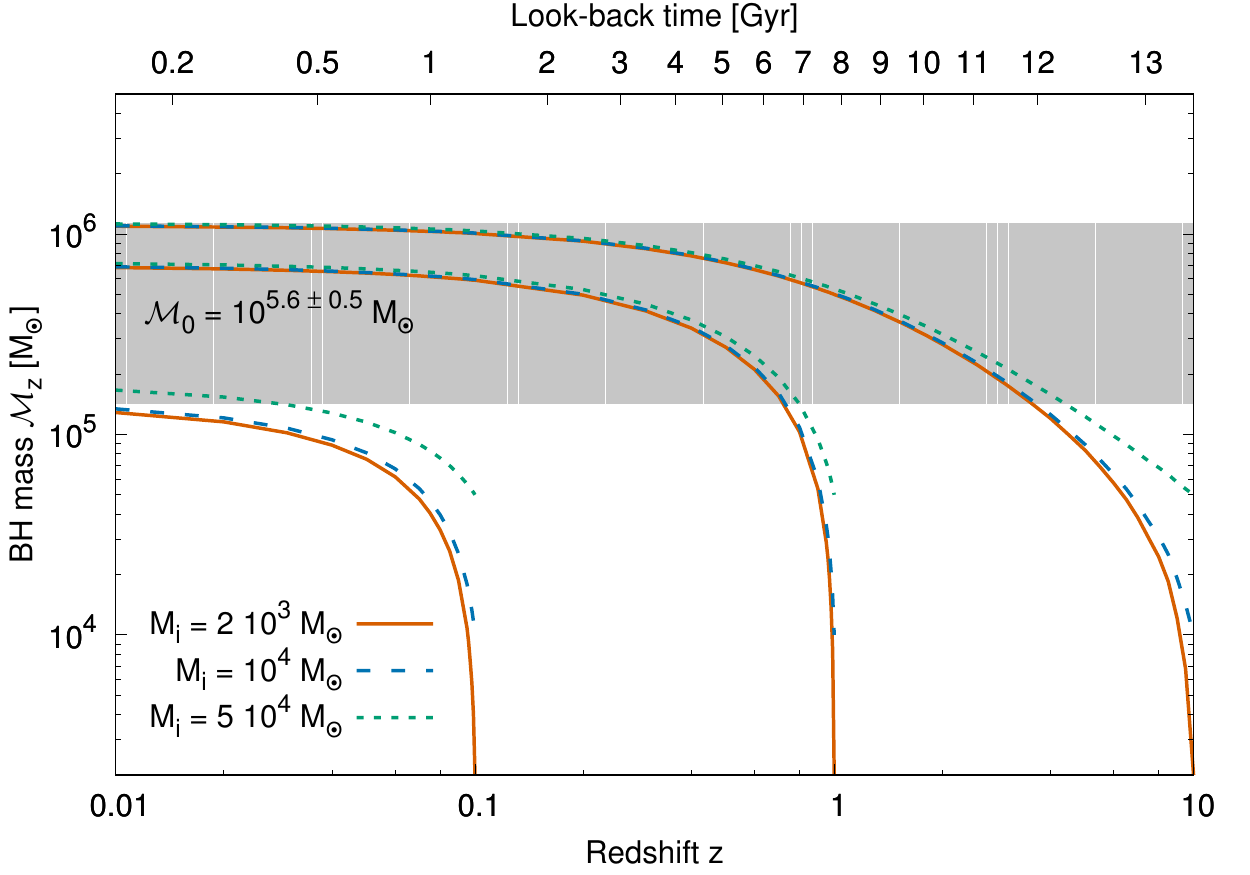}
\caption{
\label{f:Mz3zi}
  \textbf{Cosmological growth of the minimal black hole mass $\Mz$ as function
    of redshift.}  The evolution of the minimal black hole mass
  (equation~\eqref{e:asymptote}) is plotted for several values of the black
  hole seed mass $M_{i}$ at different formation redshifts $z_{i}$
  (\mbox{look-back} times\cite{ben+14} $t_{i}$), assuming the empirical $\Msig$
  relation\protect\cite{bos16} without scatter, an ${\alpha=7/4}$ cusp of Solar
  type stars and an accreted mass fraction ${f_{a}=3/8}$. The convergence to
  evolution that is independent of initial mass is rapid in redshift. The
  earlier the formation, the higher is $\Mzo$. The range ${z_{i}=0.1-10}$
  translates to a lower mass limit on present-day MBHs
  ${\Mzo\sim10^{(5.6\pm0.5)\Mo}}$ (gray band).}
\end{figure}

The solution for ${b<1}$ in the ${t\gg t_{\infty}\!=\!{(M_{i}/\Ms)}^{1-b}\!/\!(|1-b|f_{a}\Gs\!)}$
limit (equation~\eqref{e:tinf}), ${\Mbh(t)\simeq{[(1-b)f_{a}\Gs t]}^{1/(1-b)}\Ms\!\equiv\!M_\bullet^\star(t)}$,
is independent of $M_{i}$. Because ${t_{\infty}\ll t_{H}}$ for ${M_{i}\lesssim 10^{5}\Mo}$,
all seeds reach the same mass scale after $\mathcal{O}(t_{H})$ (Fig.~\ref{f:Mz3zi}). 

Consider next a realistic black hole that grows also by gas accretion and/or
mergers, ${\dot{M}_{\bullet}=\dot{M}_{\bullet}^{\star}+\dot{M}_{\bullet}^{+}}$,
where ${\dot{M}_{\bullet}^{+}\ge0}$ is the accretion rate by the non-stellar
channels. The full growth equation is 
\begin{equation}
\dot{M}_{\bullet}=\dot{M}_{\bullet}^{\star}+\dot{M}_{\bullet}^{+}\ge f_{a}\Ms\Gamma_{\star}{(\Mbh/\Ms)}^{b}\,,\,\,\,\,\,\Mbh(0)=M_{i}\,.\label{e:mutot}
\end{equation}

The solution $M_{\bullet}^{\star}(t)$ of the stars-only growth (equation
\eqref{e:dMbhdt}) then provides a lower limit\emph{ }on\emph{ }the
actual mass $\Mbh(t)$ of the growing black hole. Note that $M_{\bullet}^{\star}$
is not necessarily a lower limit on the actual stellar mass contribution
$M^{\star}$ to $\Mbh$: ${M_{\bullet}^{\star}\le M^{\star}\le\Mbh}$
for ${b\ge0}$, but ${M^{\star}<M_{\bullet}^{\star}\le\Mbh}$ for ${b<0}$.
The universal minimal mass scale of a central black hole at $t_{H}$, and
the corresponding minimal velocity dispersion scale are then
\begin{equation}
\label{e:Mh}
\small{\!\!\Mzo\!=\!M_{\bullet}^{\star}(t_{H}\!)\!=\!{[(1\!-\!b)f_{a}\Gs t_{H}\!]}^{1\!/\!(1\!-\!b)}\!\Ms,\,\,\Szo\!=\!{(\Mzo/M_{s}\!)}^{1\!/\!\beta}\!\sigma_{s}},
\end{equation}
with the index $b$ for the ${\Mbh<M_{c}}$ branch of equation~\eqref{e:TDEb74}.
This implies that galaxies with ${\ss<\Szo}$ do not have a central
black hole, or have formed it only recently, for otherwise the coevolution
of the black hole and nucleus over $t_{H}$ would have driven $\ss$ to a
much larger present-day value.

Assuming the universal $\Msig$ relation and Solar type stars, the
range in $\Mzo$ values is due to the variety and uncertainty in the
properties of galactic nuclei and of tidal disruption, and to the
intrinsic scatter in the relation. $\Mzo$ increases with the index
$\alpha$ of the stellar cusp (${\ns\propto r^{-\alpha}}$) inside $\rh$,
and with the accreted mass fraction $f_{a}$, from ${10^{5}\Mo}$
for the shallowest possible cusp of unbound stars (${\alpha=1/2}$,
${f_{a}=1/4}$) to ${10^{6}\Mo}$ for a steep isothermal cusp (${\alpha=2}$,
${f_{a}=1/2}$). ${\Mzo\simeq10^{6}\Mo}$ for the parameters adopted
here: a dynamically relaxed cusp\cite{bar+13} where ${\alpha=7/4}$,
and an accreted mass fraction ${f_{a}=3/8}$.

The scatter around the $\Msig$ relation likely reflects intrinsic
physical differences between galaxies beyond the measurement errors
on the $\Msig$ parameters\cite{bos16}. This induces roughly Gaussian
probability distributions for $\Mzo$ and $\Szo$: ${\Mzo=(1.1\pm0.8)\times10^{6}\Mo}$
and ${\Szo=79\pm35\,\mathrm{km\,s}^{-1}}$ $(1\sigma)$. The lowest-$\ss$
galaxies known to harbor active galactic nuclei\cite{xia+11} (and hence black holes, with
estimated masses ${10^{5}\lesssim\Mbh\lesssim10^{6}\Mo}$), have ${\ss\sim30-40\,\mathrm{km\,s}^{-1}}$.
This corresponds to the 5\% lower limits ${\Szo\lesssim40\,\mathrm{km\,s}^{-1}}$
and ${\Mzo\lesssim3\times10^{5}\Mo}$, which we adopt here as representative
lower limits. Lighter black holes are much rarer yet: e.g. ${\Mbh\le10^{4}\Mo}$
is below the $0.0001\%$ limit. 

The agreement ${\Mzo\lesssim\min(\Mbh^{\mathrm{obs}})\sim10^{6}\Mo}$
follows directly from basic local physics (tidal disruption and loss-cone
dynamics) and empirical global properties of the Universe (its age
and a universal $\Msig$ relation). Our derivation of $\Mzo$ rests
on four assumptions. (1) There is effective accretion of tidally disrupted
stars ($f_{a}$ is a few  ${\!\times\,0.1}$)\cite{aya+00}. (2) Most
black hole seeds were formed early, at a \mbox{look-back} time ${t_{i}\sim{\mathcal{O}}(t_{H})}$.
(3) Black hole growth is not typically mass or density limited; that is, the
growing black hole is embedded in a stellar system with ${M_{\mathrm{sys}}\gg\rh^{3}\Ms\ns>\Mbh}$
for a substantial fraction of $t_{H}$. (4) The boundary conditions
at $\rh$ are set by the universal $\Msig$ relation at all times.

An early start for black hole seeds, and the requirement that a system that
can form and retain a seed black hole should be dense and massive enough,
are both physically plausible and possibly even essential\cite{vol12}.
Such a system can be approximated as embedded in an isothermal density
distribution, and is dynamically relaxed (see \SI). Furthermore,
the accretion rate of stars in a system with $\Ns$ stars inside $\rh$,
${\mathrm{d}\Mbh/\mathrm{dt}\simeq f_{a}\Ms\Ns/(\log(1/j_{lc})T_{R})}$
(equation~\eqref{e:Rpexact}), is slow enough to allow it to remain near
equilibrium as it grows, as the timescale for growth by order of
the stellar mass, ${(\mathrm{d}\Mbh/\mathrm{dt})/(\Ms N_{\star})}$,
is longer by a factor ${\log(1/j_{lc})/f_{a}\gg1}$ than the timescale
to return to steady-state, ${T_{SS}\simeq T_{R}/4}$~(ref.~\cite{bar+13}).
The least secure assumption is that a universal $\Msig$ relation
holds near its present-day value as the black hole grows. However, this is
broadly consistent with observations of active galactic nuclei\cite{she+15,sal+13} up to ${z\sim1}$,
and with simulations of large scale structure evolution\cite{sij+15,tay+16} up to ${z\sim4}$.

Mergers between two cental black holes increase the black hole mass, but also
affect the dynamics around it. Mergers initially enhance the TDE
rate\cite{che+11}, but later they may strongly suppress it by ejecting the
cusp. However, steady-state is quickly re-established around IMBHs,
${T_{SS}(\Mbh\!\lesssim\!10^{5}\Mo)<\mathcal{O}(10^{8}\,\mathrm{yr})}$~(ref.~\cite{bar+13}).
Additional growth channels thus only increase present-day black hole masses,
and reinforce the conclusion that central black holes with ${\Mbh<\Mzo}$ are
rare.

Figure~\ref{f:Mz3zi} shows the evolution of the lower mass limit
$\Mz$, which increases rapidly with decreasing redshift to its present-day
value $\Mzo$. Present-day IMBHs may exist in recently formed systems,
in mass-limited ones (for example globular clusters with ${M<10^{6}\Mo}$),
in sub-galactic systems where the $\Msig$ relation need not apply
(such as globular clusters or super star clusters), or in very low density
galaxies (such as cored dwarfs\cite{wal+09}). However, it is unlikely
that such systems can form a black hole seed to start with\cite{mil+04b,vol12},
and therefore the black hole occupation fraction there is probably low. Candidate
IMBHs have been reported in globular clusters and dwarf galaxies,
including recently\cite{bal+15,kiz+17}, but the evidence remains
inconclusive.

Early TDE-driven growth and the suppression of the cosmic black hole mass
function below $\Mzo$ have implications for black hole seed evolution, for
the cosmic rates and properties of TDEs\cite{ree88}, and for gravitational waves
(GWs) from IMBH-IMBH mergers and intermediate-mass-ratio inspirals into IMBHs. We conclude by listing these briefly.

A high rate of TDEs can allow black hole seeds to continue growing despite
the ejection of the ambient gas by supernovae feedback\cite{dub+15}.
The lack of IMBHs at low redshifts means that electromagnetic searches
will have to reach very deep to detect TDEs from IMBHs (jetted TDEs
may provide an opportunity\cite{fia+16}). The prospects of detecting
exotic processes related to IMBHs, such as tidal detonations of white
dwarfs in the steep tidal field of a low-mass black hole\cite{ros+09},
will be low. The mean observed TDE rate per galaxy, ${\Gamma\sim10^{-5}\,\mathrm{yr^{-1}\,gal^{-1}}}$,
is much lower than predicted rates\cite{sto+16}. A dearth of black holes
below $\Mzo$ may partially resolve the rate discrepancy. 

IMBHs produce GWs by intermediate-mass-ratio inspirals and by IMBH
mergers. Detection of intermediate-mass-ratio inspirals by planned space-borne
GW observatories\cite{ama+13,yag13} is limited to redshifts below a few
${\!\times\,0.1}$, and is therefore unlikely. However, IMBH mergers can be
detected to very high redshifts. A GW search for IMBH mergers and
intermediate-mass-ratio inspirals can reveal the formation history of black
holes. We predict that black hole seeds are driven early on to higher mass by
the accretion of stars, and therefore IMBHs are rare in the present-day
Universe, but will be found near their high formation redshifts.

\renewcommand\theequation{\arabic{equation}}
\setcounter{figure}{0}

\begin{methods}

  \vspace{12pt} This supplement summarizes results from loss-cone theory used
  to derive the equation for black hole growth by stellar disruptions, and
  discusses the properties of its solutions. We first present, without
  derivation, a recipe for the approximate power-law growth rate equation
  (equation~\eqref{e:dMbhdt}), which has the advantage of leading to simple
  analytic results. We then comment on the general properties of its solutions,
  to clarify under what circumstances, and to what extent can simple
  dimensional analysis be used to estimate the minimal mass limit $\Mzo$. We
  then describe how the intrinsic scatter in the $\Msig$ relation is propagated
  through the growth equation to obtain the probability distributions for the
  lower limits $\Mzo$ and $\Szo$. Finally, we present for completeness and
  reproducibility an outline of the derivation of the full growth rate equation
  (used to verify our approximations, see Figure~\ref{f:Rprc}) and of its
  power-law approximation.

\subsection{Approximate power-law black hole growth rate equation.}

We focus here on a steady-state stellar system around a black
hole\cite{bah+76}, which has a density cusp ${\ns\propto r^{-\alpha}}$ with
${\alpha=7/4}$ inside the radius of influence
${\rh=GM_{\bullet}/\sigma_{\star}^{2}}$. We further assume that the cusp is
embedded in an external isothermal stellar distribution,
${\rho(r)=\sigma_{\star}^{2}/(2\pi Gr^{2})}$, so that the
stellar mass enclosed inside $\rh$ is twice the black hole
mass\cite{mer13}. Under the assumption of a universal $\Msig$ relation
${\Mbh={M_{s}(\ss/\sigma_{s})}^{\beta}}$, the dynamics leading to tidal
disruption are characterized by a critical mass scale
${M_{c}\sim10^{6}\Mo}$. Tidal disruptions are dominated by stars originating
from ${\sim \rh(\Mbh)}$ for ${\Mbh\ge M_{c}}$, and by stars originating from an
inner critical radius ${r_{c}(\Mbh)<\rh(\Mbh)}$ for ${\Mbh<M_{c}}$ (see below
for more details)\cite{lig+77}. The TDE rate is well-approximated by a broken
power-law (Figure~\ref{f:Rprc})
\begin{equation}
\Gamma\simeq\Gamma_{\star}{(\Mbh/\Ms)}^{b}\,,\label{e:TDEratePL}
\end{equation}
whose index $b$ changes across $M_{c}$, which is given by (see equations~(\ref{e:Rp}\textendash\ref{e:Gammas}) for the general case),
\begin{equation}
M_{c}\simeq M_{\star}{\left(\frac{16}{5s^{2}}\right)}^{3\beta/(6+\beta)}\,,\label{e:TDEQc}
\end{equation}
in terms of the dimensionless velocity dispersion scale ${s={(M_{s}/M_{\star})}^{-1/\beta}\sigma_{s}/v_{\star}}$,
where ${v_{\star}^{2}=G\Ms/\Rs}$ and $M_{\star}$ and $R_{\star}$
are the mass and radius of a typical star in the system, and where
we approximated the logarithmic term (equation~\eqref{e:Llc}) appearing in the general expressions
by a typical value $\Lambda_{lc}=2$. The index $b$ is (see
equation~\eqref{e:bRp} for the general case), 
\begin{align}
b & =\begin{cases}
(105-23\beta)/27\beta & M_{\bullet}\le M_{c}\\
(3-\beta)/\beta & M_{\bullet}>M_{c}
\end{cases}\,.\label{e:TDEb74}
\end{align}

Note that ${b\ll1}$ for the empirically determined range of the $\Msig$
relation index\cite{gra16,kor+13}, ${4\lesssim\beta<6}$. Defining
${t_{\star}=\sqrt{R_{\star}^{3}/G\Ms}}$, the rate factor is 
\begin{equation}
\Gamma_{\star}\simeq\frac{5/4}{t_{\star}}s^{40/9}\begin{cases}
1 & M_{\bullet}\le M_{c}\\
{(M_{c}/M_{s})}^{4(6+\beta)/27\beta} & M_{\bullet}>M_{c}
\end{cases}\,.\label{eq:TDEratesPL}
\end{equation}

To summarize, the approximate power-law TDE rate for a black hole with mass
$M_{\bullet}$ is calculated as follows. (1) Calculate the critical
mass $M_{c}$ (equation~\eqref{e:TDEQc}). (2) Calculate the power-law index
$b$ (equation~\eqref{e:TDEb74}) according to the low or high mass branch,
depending on $M_{\bullet}$, and similarly calculate the rate factor
$\Gamma_{\star}$ (equation~\eqref{eq:TDEratesPL}). (3) Use equation~\eqref{e:TDEratePL}
to obtain the TDE rate from $\Gamma_{\star}$ and $b$.

\subsection{Properties of the black hole growth solutions.}

The general solution of the growth equation (equation~\eqref{e:TDEratePL})
with the initial condition ${M_{\bullet}(t=0) = M_{i}}$ is
\begin{equation}
M_{\bullet}(t)=\begin{cases}
M_{\star}{[{(M_{i}/M_{\star})}^{1-b}+(1-b)t/t_{a}]}^{1/(1-b)} & b\ne1\\
M_{i}e^{t/t_{a}} & b=1
\end{cases}\,,\label{e:PLsol}
\end{equation}
where ${t_{a}={(f_{a}\Gs)}^{-1}}$ is the accretion timescale.
The growth solution has three branches. The solution for ${b=1}$ diverges
exponentially to infinity in infinite time. When ${b>1}$, $\Mbh$ diverges
on a finite timescale
\begin{equation}
t_{\infty}=t_{a}{(M_{i}/M_{\star})}^{1-b}/|1-b|\,,\label{e:tinf}
\end{equation}
and is supra-exponential. The ${b<1}$ branch is sub-exponential and
diverges slowly as a power-law.

Exponential growth describes, for example, radiation pressure-regulated
accretion of gas at the Eddington limit. Supra-exponential growth
describes Hoyle-Lyttleton wind accretion\cite{hoy+39}, spherical
Bondi accretion\cite{bon52}, or their generalization of accretion
on an accelerating black hole\cite{ale+14b}. Sub-exponential growth,
which is the relevant case for tidal accretion with the universal
$\Msig$ relation, at ${t\gg t_{\infty}}$ asymptotically approaches a power law
that is independent of  seed mass $M_{i}$
\begin{equation}
M_{\bullet}(t)/M_{\star} \simeq {[(1-b)t/t_{a}]}^{1/(1-b)}\,.\label{e:asymptote}
\end{equation}

There are two mass scales in the growth equation: the initial mass $M_{i}$, and
the natural mass scale\cite{mur+91,fre+02} $M^{\mathrm{TDE}}$, obtained by
solving
${M^{\mathrm{TDE}}/M_{\star}=f_{a}\Gamma
  t_{H}=f_{a}\Gamma_{\star}t_{H}{(M^{\mathrm{TDE}}/M_{\star})}^{b}}$ with
${t_{H}=10^{10}\,\mathrm{yr}}$. This mass scale was used in past
studies\cite{mag+99,mer13,sto+16} to estimate $M_{\bullet}(t_{H})$. It is instructive to analyze the role of $M_{\bullet}^{\mathrm{TDE}}$ in the growth solutions, and identify when it can
provide a relevant estimate for the black hole mass.

The exponential and supra-exponential solutions (${b\ge1}$) are functions of
$M_{i}$ on all timescales, and $M_{\bullet}^{\mathrm{TDE}}$ plays there a role
related to the exponential or divergence timescales. Because these solutions
diverge, the black hole mass at any finite time is generally unrelated to
either $M_{i}$ or $M_{\bullet}^\mathrm{TDE}$. The asymptotic sub-exponential
solution (${b<1}$) can be written as
${M_{\bullet}(t_{H})=M_{\bullet}^\mathrm{TDE}{(1-b)}^{1/(1-b)}}$.
In this case, $M_{\bullet}^{\mathrm{TDE}}$ provides a reasonable approximation
for $M_{\bullet}$ as long as ${|b|\ll1}$. This is the indeed case for the
empirical universal $\Msig$ relation, where
${b(\alpha=7/4,\beta=5.40)\simeq-0.125}$. However, other combinations of cusp
and $\Msig$ indices can lead to arbitrarily large disparities: for example
${b(\alpha=3/2,\beta=3)\simeq0.6}$, results in
${M_{\bullet}\simeq0.1M_{\bullet}^{\mathrm{TDE}}}$.

It should be emphasized that the solution branch that describes the black hole
growth is not determined solely by the assumed growth channel~\textemdash~tidal
disruptions in this case~\textemdash~but also by the choice of boundary
conditions, which here are determined by an empirical relation. Other possible
values of cusp and $\Msig$ indices would imply very different relations between
$M_{\bullet}$ and $M_{\bullet}^{\mathrm{TDE}}$. For example, the transition to
the exponential and supra-exponential solutions (${b\ge1}$) occurs for
${\beta\le2.1}$ (for ${\alpha=7/4}$) or for ${\beta\le3}$ (for
${\alpha=1/2}$). Therefore, it is not generally true that
$M_{\bullet}^{\mathrm{TDE}}$ estimates the black hole mass. Its relevance
depends on the specific solution and on the adopted boundary conditions, and
cannot be assumed a priori.

\subsection{Intrinsic~$\Msig$ scatter and distribution of lower limits.}

The observed intrinsic scatter in the $\Msig$ relation at ${z\simeq0}$,
with r.m.s $\delta_{\epsilon}$, can be interpreted as reflecting a
variance in the initial conditions of individual galaxies at their
formation, a Hubble time $t_{H}$ ago, or a variance that developed gradually
over their individual evolutionary histories and reached the observed
rms value at $t_{H}$. 

We assume that the estimation errors in the parameters $\alpha$ and $\beta$ of
the $\Msig$ relation,
{${\log(\!\Mbh\!/\!\Ms\!)\! =\! (\bar{\alpha}\!\pm\!\delta_{\alpha}\!)\! +\!
  (\bar{\beta}\!\pm\!\delta_{\beta}\!)\!\log(\sigma\!/\!\sigma_{s}\!)\!\pm\!\delta_{\epsilon}}$},
can be approximated by a correlated bi-Gaussian distribution,
${(\alpha,\beta)\sim
  G_{2}(\bar{\alpha},\delta_{a},\bar{\beta},\delta_{\beta},\rho_{\alpha\beta})}$,
where for an arbitrarily chosen low reference velocity dispersion
${\sigma_{s}\ll\sigma}$, the correlation coefficient
${\rho_{\alpha\beta}\to-1}$~(ref.~\cite{tre+02}), and that the intrinsic scatter
$\epsilon$ is drawn from a Gaussian distribution,
${\epsilon\sim G(0,\delta_{\epsilon})}$.

We approximate the evolution of the scatter by assuming $n_{t}$ discrete time
steps of duration ${\Delta t=t_{H}/n_{t}}$, where the accumulated change in
$\alpha$ due to scatter, $\Delta\epsilon$, is modified in a random walk fashion
by ${\Delta\epsilon\to\Delta\epsilon+\epsilon/\sqrt{n_{t}}}$. We then evolve
the black hole mass over time $\Delta t$ by the growth equation
(equation~\eqref{e:PLsol}), and repeat until ${t=t_{H}}$. The joint and
marginal probability distributions for $\Mzo$ and $\Szo$ at $t_{H}$ are
obtained by Monte Carlo simulations over randomly drawn values of $\alpha$ and
$\beta$.

The limit ${n_{t}=1}$ corresponds to scatter that is determined by
the galaxy's initial conditions, whereas ${n_{t}\gg1}$ corresponds
to scatter that is determined by the galaxy's evolution. We find that
the probability distributions for $\Mzo$ and $\Szo$ do not depend
strongly on the choice of $n_{t}$, and that they converge rapidly
for ${n_{t}>3}$ to an asymptotic form. The values quoted in this study,
$5\%$ lower limits of ${\Mzo=2.8\times10^{5}\Mo}$ and ${\Szo=38\,\mathrm{km\,s}^{-1}}$,
correspond to the asymptotic evolutionary scatter case ($n_{t}=5$),
whereas the initial scatter case ($n_{t}=1)$ differs only slightly,
with $5\%$ lower limits of $\Mzo=1.9\times10^{5}\Mo$ and $\Szo=36\,\mathrm{km\,s}^{-1}$.

\subsection{Full black hole growth rate equation.}

The tidal disruption (plunge) rate can be approximated by the flux
of stars into the black hole from from the boundary between the inner region,
where stars slowly diffuse into the loss-cone (the empty loss-cone)
and the outer region, where stellar scattering is strong enough that
the loss-cone is effectively full (the full loss-cone)\cite{lig+77}.
The boundary is at a critical radius, $a_{c}$, that satisfies 
\begin{equation}
q=\frac{{[J_{c}(a_{c})/J_{lc}]}^{2}P(a_{c})}{\log(J_{c}(a_{c})/J_{lc})T_{R}(a_{c})}=1\,,\label{e:qfactor}
\end{equation}
where ${P=2\pi\sqrt{a^{3}/GM_{\bullet}}}$ is the orbital period, ${J_{c}=\sqrt{GM_{\bullet}a}}$
is the circular angular momentum at $a$, $J_{lc}$ is angular momentum
of the loss-cone (${J_{lc}\simeq\sqrt{2G\Mbh Q^{1/3}\Rs}}$ for tidal
disruption, so ${J_{c}/J_{lc}\simeq\sqrt{(a/\Rs)/2Q^{1/3}}}$, where
${Q=M_{\bullet}/M_{\star}}$). $T_{R}$ is the 2-body (non-resonant)
relaxation time\cite{bar+13}, 
\begin{equation}
T_{R}(a)=\frac{5}{8}\frac{Q^{2}P(a)}{\Ns(a)\log(Q)}\,,\label{e:TR}
\end{equation}
where ${\Ns(a)=\mu_{h}Q{(a/\rh)}^{3-\alpha}}$ is the number of stars
enclosed in $r$, and ${\rh=\eta_{h}GM_{\bullet}/\ss^{2}}$ is the
radius of influence. The numeric prefactors are conventionally assumed
to be ${\mu_{h}=2}$ and ${\eta_{h}=1}$. We further assume that ${\alpha<9/4}$.

The exact solution for the critical radius can be written by the implicit
equation
\begin{align}
a_{c}/\rh  
  = {} 
  &
{(A_{c}\sigma_{\star}^{2}/v_{\star}^{2})}^{1/(4-\alpha)}Q^{1/(12-3\alpha)}
    \nonumber \\
  = {} 
  & {(A_{c}s^{2})}^{1/(4-\alpha)}Q^{(\beta+6)/\beta(12-3\alpha)}\,,\label{e:qexact}
\end{align}
where ${A_{c}=5/(4\mu_{h}\Lambda_{lc}\eta_{h})}$ and
\begin{equation}
\label{e:Llc}
\Lambda_{lc}(Q,a)=\log Q/\log(J_{c}(a)/J_{lc})\,,
\end{equation}
where the last equality in equation~\eqref{e:qexact} assumes the $\Msig$
relation in terms of the dimensionless velocity dispersion scale
${s={(M_{s}/M_{\star})}^{-1/\beta}\sigma_{s}/v_{\star}}$,
where ${v_{\star}=\sqrt{GM_{\star}/R_{\star}}}$.

When ${a_{c}>\rh}$, the rate is estimated\cite{sye+99} at $\rh$.
The transition occurs above a critical black hole mass such that ${a_{c}(M_{c})=\rh(M_{c})}$,
\begin{equation}
M_{c}  =  M_{\star}{(A_{c}\sigma_{\star}^{2}/v_{\star}^{2})}^{-3}=\Ms{(A_{c}s^{2})}^{-3\beta/(6+\beta)}\,,\label{eq:Mc}
\end{equation}
which is independent of $\alpha$ and almost independent of $\beta$.
 The plunge rate can then be conservatively approximated by the empty
loss-cone rate at ${a_{e}=\min(a_{c},\rh)}$,
\begin{equation}
\Gamma\simeq\frac{\Ns(a_{e})}{\log(J_{c}(a_{e})/J_{lc})T_{R}(a_{e})}=\frac{8}{5}\frac{\Lambda_{lc}\mu_{h}^{2}}{P(\rh)}{\left(\frac{a_{e}}{\rh}\right)}^{(9-4\alpha)/2}\,.\label{e:Rpexact}
\end{equation}
The actual rate, including the contribution from the full loss-cone
regime, can be up to twice as high as this as this\cite{sye+99}.

Using ${\rh=\eta_{h}GM_\bullet/\sigma_{\star}^{2}}$, the plunge rate can
be represented as
\begin{equation}
\small{
\Gamma^{\Lambda}\!=\!\frac{1}{\pi t_{\star}}\!\begin{cases}
\gamma_{c}^{\Lambda}Q^{(2\alpha-15)/6(4-\alpha)}{\left(\!\frac{\ss}{v_{\star}}\!\right)}^{7(3-\alpha)/(4-\alpha)} & \!\!M_{\bullet}\!\le\! M_{c}\\
\gamma_{h}^\Lambda Q^{-1}{\left(\!\frac{\ss}{v_{\star}}\!\right)}^3 & \!\!M_{\bullet}\!>\!M_{c}
\end{cases}},
\end{equation}
where ${t_\star=\sqrt{R_{\star}^{3}/GM_{\star}}}$,
\begin{equation}
\gamma_{c}^\Lambda =
{\left(\frac{4\Lambda_{lc}}{5}\right)}^{(2\alpha-1)/(8-2\alpha)}
\!
{\left(\frac{\mu_{h}}{\eta_{h}^{(3-\alpha)}}\right)}^{7/(8-2\alpha)}\,,
\end{equation}
and
\begin{equation}
\gamma_{h}^{\Lambda}=\frac{4\mu_{h}^{2}\Lambda_{lc}}{5\eta_{h}^{3/2}}\,.
\end{equation}
When $\ss$ is given by the $\Msig$ relation, the rate can be expressed
as
\begin{equation}
\Gamma=\Gamma_{\star}^{\Lambda}Q^{b}\,,\label{e:Rp}
\end{equation}
where
\begin{equation}
\label{e:Gammas}
\small{
\Gamma_{\star}^{\Lambda}
\!\!=\!\!
\frac{\gamma_{c}^{\Lambda}}{\pi t_{\star}}
s^{7(3-\alpha)/(4-\alpha)}\!\!
\begin{cases}
  1 & \!\!\!M_{\bullet}\!\le\! M_{c}\\  \!\!{\left(\!\frac{M_{c}}{M_{\star}}\!\right)}^{\!(6+\beta)(9-4\alpha)/6(4-\alpha)\beta} & \!\!\!M_{\bullet}\!>\!M_{c}
\end{cases}}.
\end{equation}

The notation $\Gamma_{\star}^{\Lambda}$ denotes the weak functional dependence
on $Q$ via the logarithmic term ${\Lambda_{lc}\simeq2}$. The index is
\begin{equation}
\small{
b\!=\!\begin{cases}
7(3-\alpha)/\beta(4-\alpha)-(15/2-\alpha)/3(4-\alpha) & a_{c}\le \rh\\
(3-\beta)/\beta & a_{c}>\rh
\end{cases}}.
\label{e:bRp}
\end{equation}

A simpler power-law approximation can be obtained by choosing this typical
value for the logarithmic term ${\Lambda_{lc}\simeq2}$. Then,
${\Gamma\simeq\Gamma_{\star}Q^{b}}$ (see above), where the normalization
${\Gamma_{\star}=\Gamma_{\star}^\Lambda(\Lambda_{lc}=2)}$ is not a
function of $Q$.

Figure~\ref{f:Rprc} shows the TDE rates for ${\mu_{h}=2}$, ${\eta_{h}=1}$ in
the two dynamical regimes ${a_{c} < \rh}$ and ${a_{c} > \rh}$ (${Q < Q_{c}}$
and ${Q > Q_{c}}$). At the lower mass end, the mean mass accretion rate may
rise above the Eddington rate. For example, for ${\beta=5}$,
${L_{TD}/L_{E}\lesssim6}$ at ${\Mbh=10^{3}\Mo}$, but falls below
${L_{TD}/L_{E}=1}$ for ${\Mbh>5\times 10^{3}\Mo}$. Depending on the exact
value of the logarithmic slope of the $\Msig$ relation, $\beta$, the TDE rate
on the low-mass branch, can either rise or fall with $\Mbh$.
\end{methods}
\begin{addendum}

\item[Data availability statement]

  The numeric results that support the plots within this paper and other
  findings of this study are available from the corresponding author upon
  reasonable request.

\end{addendum}

\begin{addendum}

\item[Acknowledgements]

  We are grateful for helpful discussions with Y. Alexander, J. Gair,
  A. Gal-Yam, J. Green, J. Guillochon, M. MacLeod, N. Neumayer, T. Piran,
  E. Rossi, A. Sesana, J. Silk, N. Stone and B. Trakhtenbrot. TA acknowledges
  support by the I-CORE Program of the Planning and Budgeting Committee and The
  Israel Science Foundation (grant No 1829/12). BB acknowledges support by NASA
  (grant NNX14AM24G) and by the NSF (grant AST-1406166).

\item[Author Contributions]

  TA and BB developed the ideas presented in this paper together and
  collaborated in its writing.

\item[Correspondence]

  Correspondence and requests for materials should be addressed to Tal
  Alexander (tal.alexander@weizmann.ac.il) or Ben Bar-Or (benbaror@ias.edu).

\item[Competing Interests] 

  The authors declare that they have no competing financial interests.

\end{addendum}
\bibliography{nat}{}
\end{document}